\documentclass[preprint,preprintnumbers,amsmath,amssymb]{revtex4}

\usepackage{graphicx}
\usepackage{dcolumn}
\usepackage{bm}

\begin{document}
\title{Nernst effect and dimensionality in the quantum limit}
\author{Zengwei Zhu$^{1,2}$, Huan Yang$^{1}$, Beno\^{\i}t Fauqu\'e$^{1}$, Yakov Kopelevich$^{3}$,\\
and Kamran Behnia$^{1}$\email{kamran.behnia@espci.fr}}
\affiliation{$^{1}$Laboratoire Photons Et Mati\`ere (UPMC-CNRS), ESPCI, 75005 Paris, France\\
$^{2}$ Department of Physics, Zhejiang University, Hangzhou 310027, China\\
$^{3}$ Instituto de Fisica ``Gleb Wataghin'', Universidade Estadual de Campinas, UNICAMP\\
13083-970 Campinas, S\~{a}o Paulo, Brazil}

\begin{abstract}
 \textbf{Nernst effect, the transverse voltage generated by a longitudinal thermal gradient in presence of magnetic field has recently emerged as a very sensitive, yet poorly understood, probe of electron organization in solids.  Here we report on an experiment on graphite, a macroscopic stack of graphene layers, which establishes a fundamental link between dimensionality of an electronic system  and its Nernst response.  In sharp contrast with  single-layer graphene, the Nernst signal sharply peaks whenever a Landau level meets the Fermi level. This points to the degrees of freedom provided by finite interlayer coupling as a source of enhanced thermoelectric response in the vicinity of the quantum limit. Since Landau quantization slices a three-dimensional Fermi surface, each intersection of a Landau level with the Fermi level modifies the Fermi surface topology. According to our results, the most prominent signature of such a topological phase transition emerges in the transverse thermoelectric response.}
\end{abstract}

\maketitle

Graphene, a single layer of carbon atoms set in a honeycomb lattice, has attracted much attention because of its uncommon properties\cite{geim}. The two-dimensional gas of massless electrons embedded in graphene displays a particular version of Quantum Hall Effect\cite{novoselov,zhang}. Its thermoelectric response has been a subject of recent experimental investigations by several groups\cite{zuev,wei,check}, who found results in agreement with the theory of thermoelectricity for a two-dimensional electron system in the quantum Hall regime\cite{jonson,oji}. On the other hand, the thermoelectric tensor of graphite\cite{brandt}, a macroscopic stack of graphene layers, has never been studied at low temperatures and high magnetic fields. Here we report on such a study uncovering the crucial role of interlayer coupling. When the field is strong enough to push the system to the quantum limit, the quantum oscillations of the Nernst response easily dwarf the oscillations seen in other transport coefficients including the Seebeck signal. Moreover, the Nernst signal sharply peaks whenever a Landau level intersects the chemical potential. Both these features are absent in graphene, but  were previously reported in bulk bismuth near the quantum limit\cite{behnia1}. The resolved Nernst peaks  emerge as signatures of a topological phase transition\cite{lifshitz}, which occurs at the intersection of Landau and Fermi level \cite{blanter}. Such a field-induced modification of the Fermi surface topology at lower Landau levels is exclusive to three dimensions, for which no adequate quantitative description of the thermoelectric response in the vicinity of the quantum limit is yet available. The results point to the configurational degrees of freedom associated with the finite dispersion of electrons along the magnetic field as the source of a huge off-diagonal thermoelectric response.

Fig. 1 presents the thermal evolution of the field dependence of the Nernst signal, S$_{xy}$ in two highly-oriented pyrolytic graphite(HOPG) samples. As the temperature decreases, quantum oscillations become sharper and their visibility extends to lower fields. In graphite, quantum oscillations of various physical properties such as resistivity (the Shubnikov-de Hass effect)\cite{soule}, magnetic susceptibility (the de Hass-van Alphen effect)\cite{williamson}, the Hall coefficient and thermopower\cite{woollam} were measured many years ago and the results are in agreement with the structure of the Fermi surface predicted by the Slonczewski-Weiss-McClure (SWM) model\cite{slonczewski,mcclure}. More recently, quantum oscillation studies employing new techniques for analysis and measurements have emerged and the degree of accuracy by which the SWM model describes the fine structure of the Fermi surface has become a subject of ongoing debate\cite{luk,schneider}. However, in the case of the Nernst effect, the only available report  is restricted to temperatures above the liquid nitrogen\cite{mills}.

The particularity of the Nernst effect as a probe of quantum oscillations at low index Landau levels is highlighted in Fig. 2. Quantum oscillations are visible in the field dependence of both diagonal and off-diagonal components of electric and thermoelectric conductivity tensors. However, only in the case of transverse thermoelectric response, the oscillating signal peaks by far the non-oscillating background. In contrast, the Shubnikov-de Haas oscillations represent a tiny fraction of the overall longitudinal resistivity. Moreover, while the longitudinal response dominates for resistivity  ($\rho_{xx} \gg \rho_{xy}$), the Nernst signal is an order of magnitude larger than the Seebeck coefficient ($S_{xy} \gg S_{xx}$). The large magnitude of the Nernst coefficient in graphite (second only to bismuth\cite{behnia3} among metals) is an expected consequence of the high mobility ($ 3\times 10^5 cm^{2} V^{-1} s^{-1}$) and the low Fermi energy (19 meV) of the system (See the supplementary information for details).

Fig. 3 compares the fine structure of $\rho_{xy}$ and S$_{xy}$ when the lowest Landau levels cross the Fermi level. The Nernst response peaks when the magnetic field attains a value corresponding to an intersection of a Landau level with the the Fermi level. The magnitude of these fields were determined by theory\cite{sugihara} and confirmed by  experiment\cite{woollam} many years ago (See supplementary information for details on the indexing of the Landau levels and a comparison of the fine structure in $S_{xy}$ and S$_{xx}$). The  peaks in S$_{xy}$  for the lowest Landau levels of the two pockets can be clearly associated with a jump or drop in $\rho_{xy}$ depending on the sign  (hole-like or electron-like) of the carriers of the pocket. Between jumps and drops, that is when the Fermi level lies between distinct Landau levels, the Hall resistivity, while not strictly flat, shows little field dependence. Presenting the same data in a different fashion in the lower panel of the same figure, one can see that at high fields and low temperatures, $\frac{\nu}{T}$ becomes constant between two successive peaks. In other words, in the low-T high-B limit, when a Landau level does not intersect the Fermi level , the Nernst response becomes both $B$-linear and $T$-linear but sharply peaks otherwise, with a singularity which becomes more pronounced with cooling. In addition to the two above-mentioned HOPG samples, we also measured a natural single crystal of graphite and found very similar results (see the supplementary information).

It is illuminating to compare these features with results very recently reported for graphene\cite{zuev,check}. The first difference is the sheer magnitude of the Nernst response. Whereas in graphene, the measurable signal at T=10 K is in the range of 10 to 20 $\mu V K^{-1}$, the Nernst signal in the same temperature range in graphite is two orders of magnitude larger and approaches  1 mV $K^{-1}$ (Fig. 1a). This difference is most probably due to a difference in electron mobility. In graphite, it can exceed $10^6 cm^2 V^{-1}s^{-1}$\cite{soule2}, two orders of magnitude larger than most currently available graphene samples, including those used in the Nernst studies. The second and more fundamental difference is the field profile of the Nernst response. In graphene, the intersection of the chemical potential and a Landau level (which can be realized either by scanning the field or modifying the gate voltage) leads to a change in the sign of S$_{xy}$. In other words, the coincidence of a Landau level with the chemical potential is concomitant with a jump in $\rho_{xy}$ and a \emph{vanishing} $S_{xy}$. In graphite, as seen above, the Nernst response \emph{attains its maximum} in the same conditions. Note that the huge electron mobility, the origin of the large low-field non-oscillating Nernst response in graphite, can neither explain the preponderance of the oscillating component, nor its particular profile.

Interestingly, the field profile of the Nernst response we find here is analogous to the one previously reported in bulk bismuth\cite{behnia1}. On the other hand, the Nernst response of graphene\cite{zuev,check} presents a functional form  reminiscent of the case of the two-dimensional electron systems realized in semiconductor heterostructures\cite{fletcher}. In both latter cases, S$_{xy}$ vanishes at the intersection of Landau and Fermi levels as predicted by the theory conceived for two dimensions\cite{jonson,oji,gusynin}.

Two features set apart graphene from both  bismuth  and bulk graphite. The most fundamental is the presence of a finite third-axis dispersion in the bulk materials. A finite interlayer coupling, no matter how small, would warp the perfect cylindrical Fermi surface of a two-dimensional monolayer. According to the SWM model, the Fermi surface of both electrons and holes in graphite are elongated ellipsoids. The precise magnitude of interlayer coupling has been a matter of debate and the Fermi surface of one type of carriers may not be closed along the c-axis. In any case, however, it cannot be a perfect cylinder. The second difference is that both bismuth and bulk graphite are compensated semi-metals hosting equal concentrations of mobile carriers of both sign. Experiments on graphene were performed on a system with a single type of carriers determined by the sign of the applied gate voltage. These two features lead to the emergence of a qualitatively different transverse thermoelectric response.

The consequences of the first difference is sketched in fig. 4, which compares the passage of successive Landau levels through the chemical potential in presence (3D, left ) or absence (2D, right) of the z-axis dispersion. In three dimensions, Landau quantization truncates the Fermi surface. The wave-vector of mobile electrons are confined to slices formed by the intersection of the Fermi-Dirac distribution and Landau spectrum. The thickness of these slices depends on temperature and disorder. As the field is swept the slices move outward and downward. The Nernst peaks are concomitant with the merger of two slices. As first noticed many years ago\cite{blanter}, such a merger is a case of electronic topological phase transition\cite{lifshitz}. Interestingly, a singularity in the Nernst response in the vicinity of a topological transition was theoretically predicted in another context\cite{livanov}. On the other hand, in the two-dimensional case, sketched in the right side of the same figure, no such topological transition occurs when the ellipsoid is replaced by a perfect cylinder. Here, when the Landau level coincides with chemical potential, the system becomes dissipative, otherwise it is gapped.

In the two-dimensional case, displacing a Landau level across the chemical potential leads to a symmetric S$_{xy}$ profile. This can be qualitatively understood in the semi-classical picture. The thermoelectric tensor $\widetilde{S}$ measures the change in conductivity tensor, $\widetilde{\sigma}$,  caused by a small shift in the chemical potential:

\begin{equation}\label{1}
\widetilde{S}= \frac{\pi^{2}}{3}\frac{k_{B}^{2}T}{e}\widetilde{\sigma}^{ -1}\frac{\partial\widetilde{\sigma}}{\partial\epsilon}|_{\epsilon=\mu}
\end{equation}

According to this equation (sometimes called the extended Mott formula), if the chemical potential happens to be at its optimal position for a maximal Hall mobility $\frac{\sigma_{xx}}{\sigma_{xy}}$, the transverse thermoelectric response, S$_{xy}$,  is expected to vanish. This simple argument can explain the qualitative profile of the  Nernst response in the 2D case, which is symmetric and vanishes at the intersection. Since when a Landau level is precisely at the chemical potential, any shift of the latter would diminish the Hall mobility. This zero response should be sandwiched between a positive and a negative peak (as sketched in Fig.4d) . The sign of these two reflect the sign of the energy derivative when one approaches the critical field from lower fields or higher fields.

The profile of the Seebeck coefficient in the 3D graphite (See Fig. S3 in the supplementary information) is not qualitatively different from the 2D one. On the other hand, the Nernst response in the 3D configuration emerges as a remarkable singularity. The clear asymmetry between configuration a and configuration c in Fig. 4 is the key to the functional shape of the Nernst response. Experimentally, S$_{xy }$ steadily increases as the two separated slices of the Fermi surface move towards each other. It peaks when they merge to form a single one and begins a sudden drop after the Landau level moves beyond the Fermi level.  The configurational entropy of the electronic states with vanishing k$_{z}$ and quantized k$_{xy}$, which emerge as this topological phase transition occurs, appears to be the source of the enhanced Nernst response exclusive to the 3D case.

The compensated nature of the bulk semi-metals under study imposes another constraint. Charge neutrality implies that any change in the density of carriers with one sign would shift the chemical potential in a way to maintain the equality between the concentration of electron and holes. A topological transition in one (say hole-like) ellipsoid thus moves the chemical potential and modifies the shape of the other.

When electrons are confined to two dimensions, the quantum oscillations of the Hall response become sharp steps in the vicinity of the quantum limit. It emerges from this study that such a confinement modifies also the quantum oscillations of the Nernst response in a remarkable way.A satisfactory theoretical understanding of the Nernst effect in bulk systems across the quantum limit is still missing. It is a remarkable irony that the least theoretically understood transport coefficient happens to be the most experimentally sensitive probe of quantum oscillations. 

Extending theses measurements to higher magnetic fields would let one probe the thermoelectric response of graphite in the extreme quantum limit. Such a study in bismuth has recently uncovered several enigmatic field scales\cite{behnia2}. One collateral conclusion of study presented here is to rule out that the Nernst anomalies seen there could originate from the two-dimensional surface states. Indeed, the field profile of the Nernst anomalies observed in bismuth clearly points to a three-dimensional origin.

We thank A. Varlamov for stimulating discussions and in particular for introducing ref.\cite{blanter} to us. This work is supported by the Agence Nationale de Recherche (ANR-08-BLAN-0121-02) as a part of DELICE project in France and by FAPESP and CNPq in Brazil. Z.Z. acknowledges a scholarship granted by China Scholarship Council.

\newpage
\begin{figure}
{\includegraphics[width=12cm]{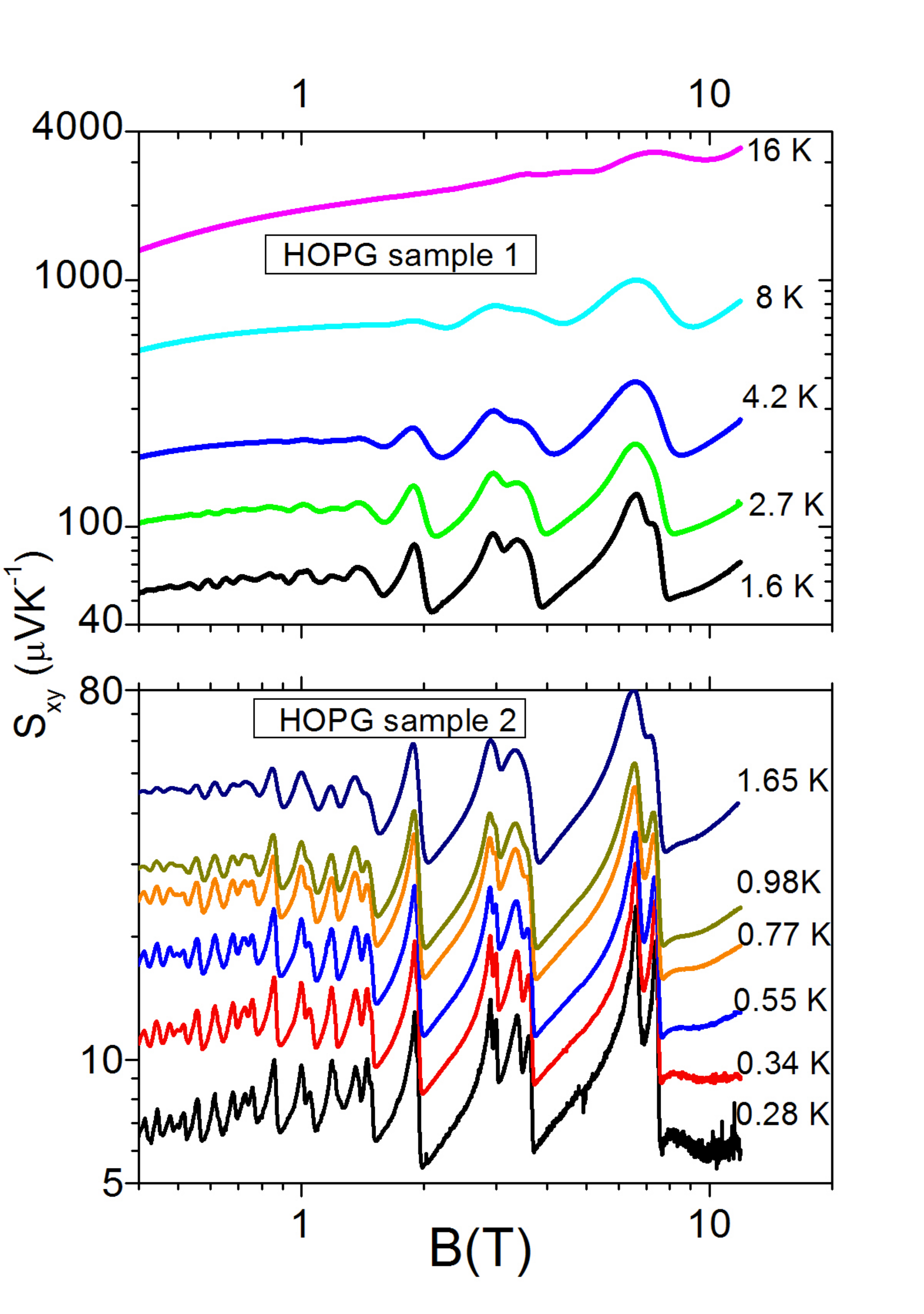}}\caption{Zhu \emph{et
al.}}
\end{figure}
\noindent {\bf Fig. 1.} The field dependence of the Nernst signal in two Highly-Oriented Pyrolytic Graphite (HOPG) samples at different temperatures. The signal was measured with an in-plane thermal gradient and a field oriented perpendicular to the layers. As the temperature is lowered, oscillations become sharper and the Zeeman splitting becomes visible.
\newpage
\begin{figure}
{\includegraphics[width=10cm]{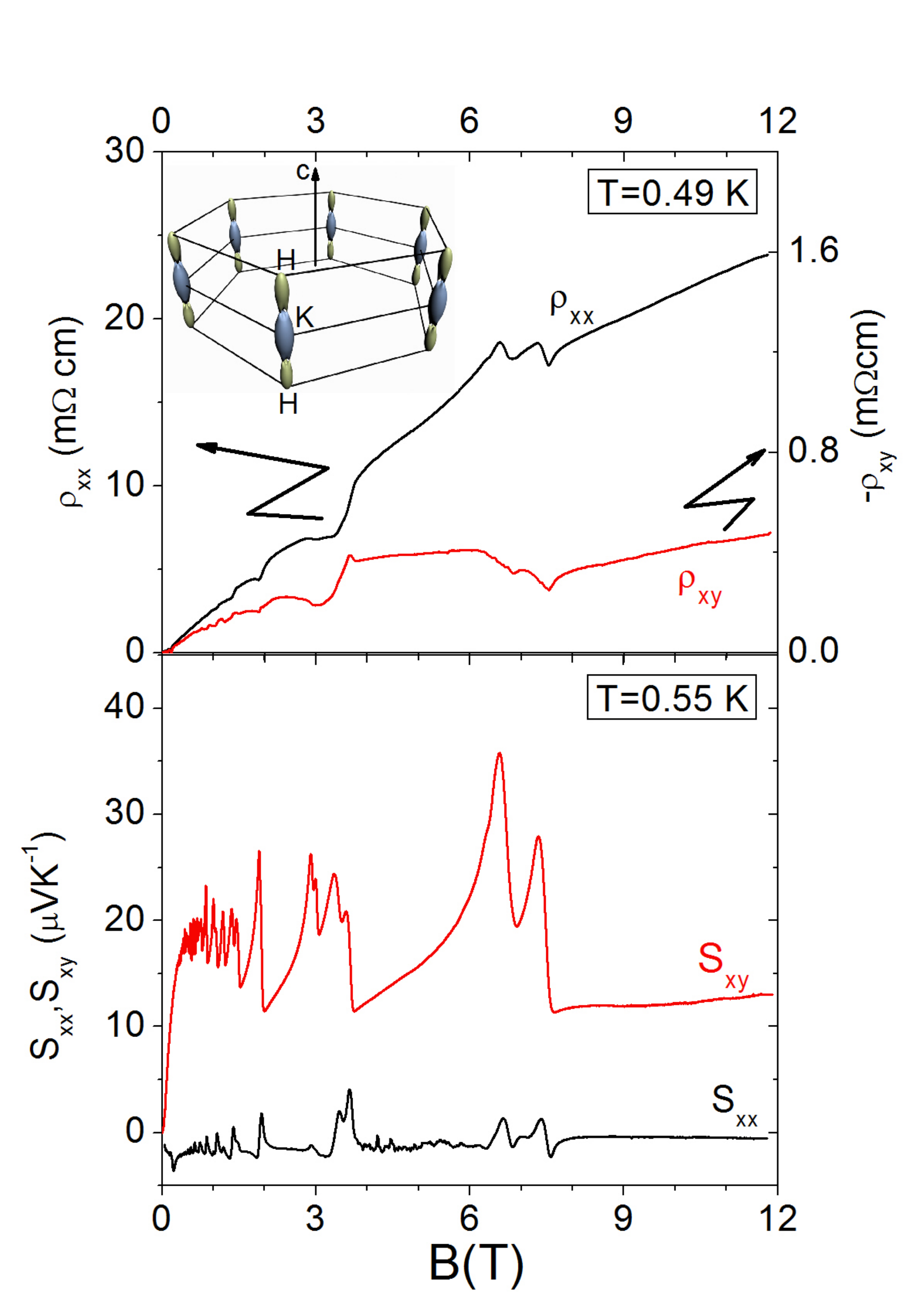}}\caption{Zhu  \emph{et
al.}}
\end{figure}
\noindent {\bf Fig. 2.} Quantum oscillations of various transport coefficients  at low temperatures. The upper panel shows longitudinal and Hall resistivity of sample 2. Shubnikov-de Haas oscillations superpose on a huge non-oscillating background. The lower panel shows the field dependence of the thermoelectric coefficients in the same sample. The Nernst response dominates by far the Seebeck coefficient and presents pronounced well-defined oscillations. The Fermi surface and the Brillouin zone of graphite are shown as an inset in the upper panel. The Fermi surface consists of 6 pairs of adjacent ellipsoid pockets (in blue and green) hosting carriers of opposite signs.
\newpage
\begin{figure}
{\includegraphics[width=10cm]{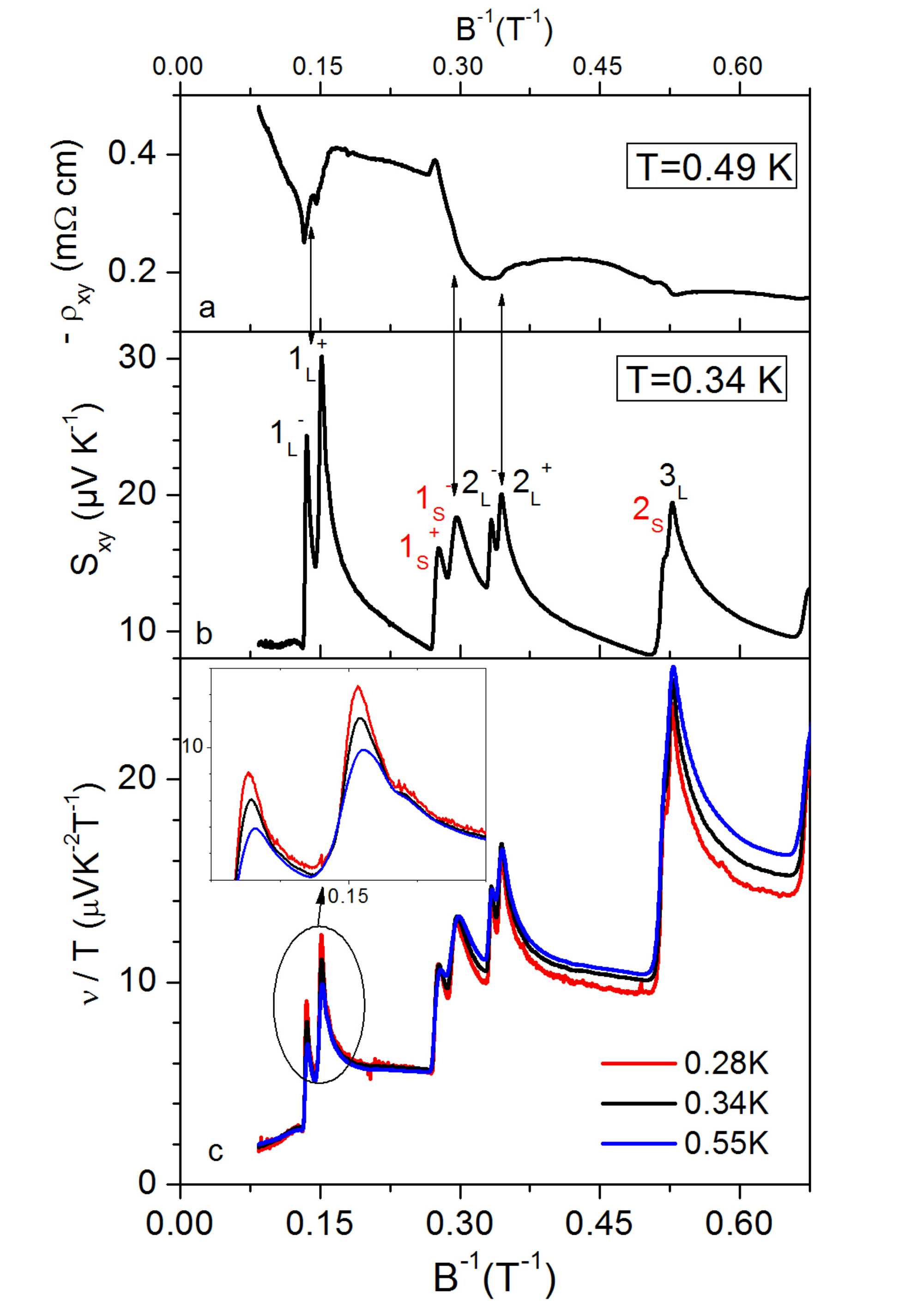}}\caption{Zhu \emph{et
al.}}
\end{figure}
\noindent {\bf Fig. 3.} The variation with inverse magnetic field of the Hall resistivity,$\rho_{xy}$ (a), the Nernst signal, S$_{xy}$ (b), and the Nernst coefficient divided by temperature, $\frac{\nu}{T} =\frac{Sxy}{B T}$, at 3 different temperatures (c). When a Landau level meets the Fermi level, the Hall resistivity presents a jump and the Nernst signal sharply peaks. At low temperature and high field, $S_{xy}$  becomes linear in field and temperature when the Fermi level is between two Landau levels. The inset zooms at  peak positions. When a Landau level is at the Fermi level,  $\frac{\nu}{T}$ increases with decreasing temperature, otherwise it attains a temperature-independent magnitude. Landau levels of the pockets are indexed in panel b.
\newpage
\begin{figure}
{\includegraphics[width=10cm]{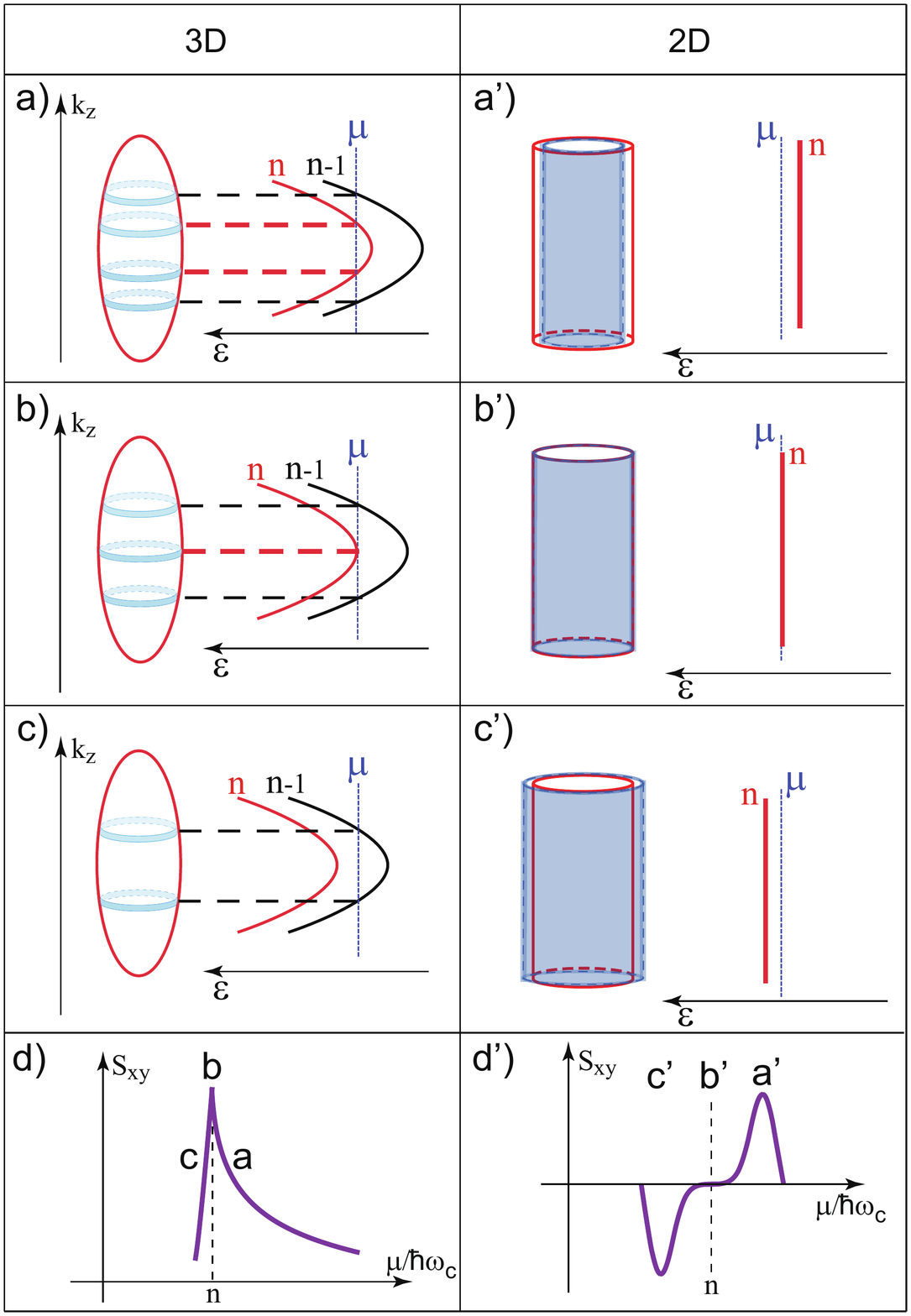}}\caption{Zhu \emph{et
al.}}
\end{figure}
\noindent {\bf Fig. 4.} A comparison of how Landau levels (LLs, indexed n, n-1) cross the chemical potential, $\mu$, as the magnetic field is swept, in  the 3D (panels a to c)  and 2D (panels a' to c') cases. In each panel the Fermi surface  with cuts of the Landau cylinders in reciprocal space is shown together with a plot of the corresponding energies. The z-axis is set by the direction of the applied magnetic field. In the 3D case , an ellipsoidal Fermi surface is truncated to slices, which are the intersections of the Fermi level and the Landau levels.  In the 2D case, the Fermi surface is a perfect cylinder and there are no slices. The Nernst signal is sketched in the panels d and d'. In 3D, S$_{xy}$ peaks when the Landau level meets the Fermi level and a topological phase transition occurs by the merger of two slices. The odd number of slices in panel b is to be contrasted with the even number of slices in panel a and c. In 2D,  S$_{xy}$ vanishes when a Landau level intersects the chemical potential. The experimentally resolved Nernst signal presents the profile of panel d in both graphite [as reported here] and bismuth\cite{behnia1}. On the other hand, in graphene\cite{zuev,wei,check} and semiconducting heterostructures\cite{fletcher} it presents the profile of panel d'.


\begin{thebibliography}{}
\bibitem{geim} A. K. Geim and K. S. Novoselov, The rise of graphene, Nature Mater. \textbf{6}, 183 (2007).
\bibitem{novoselov} K. S. Novoselov \emph{et al.}, Two-dimensional gas of massless Dirac fermions in graphene,  Nature  \textbf{438}, 197(2005).
\bibitem{zhang} Y. Zhang \emph{et al.}, Experimental observation of the quantum Hall effect
and Berry's phase in graphene, Nature \textbf{438}, 201 (2005).
\bibitem{zuev} Y. M. Zuev, W. Chang and P. Kim, Thermoelectric and Magnetothermoelectric Transport Measurements of Graphene, Phys. Rev. Lett., \textbf{102}, 096807 (2009).
\bibitem{wei} P. Wei \emph{et al.}, Anomalous Thermoelectric Transport of Dirac Particles in Graphene, Phys. Rev. Lett. \textbf{102}, 166808 (2009).
\bibitem{check} J. G. Checkelsky and N. P. Ong, The thermopower and Nernst Effect in graphene in a magnetic field, arXiv: 0812.2866.
\bibitem{jonson} M. Jonson and S. M. Grivin, Thermoelectric effect in a weakly disordered inversion layer subject to a quantizing magnetic field, Phys. Rev. B \textbf{29}, 1939 (1984).
\bibitem{oji} J. Oji, Thermomagnetic effects in two-dimensional electron systems, J. Phys. C \textbf{17}, 3059 (1984).
\bibitem{brandt} N. B. Brandt, S. M. Chudinov and Ya. G. Ponomarev, Semimetals I. Graphite and its compounds (Elsevier, Amsterdam 1988).
\bibitem{behnia1} K. Behnia, M. -A. M\'easson and Y. Kopelevich, Oscillating Nernst-Ettingshausen Effect in Bismuth across the Quantum Limit, Phys. Rev. Lett. \textbf{98}, 166602 (2007).
\bibitem{lifshitz} I. M. Lifshitz, Anomalies of electron characteristics of a metal in the high pressure region, JETP \textbf{11}, 1130 (1960).
\bibitem{blanter} Ya. M. Blanter, M. I. Kaganov, A. V. Pantsulaya and A. A. Varlamov, The theory of electronic topological transitions, Phys. Rep. \textbf{245}, 159 (1994).
\bibitem{soule} D. E. Soule, J. W. McClure and L. B. Smith, Study of the Shubnikov-de Haas Effect. Determination of the Fermi Surfaces in Graphite, Phys. Rev. \textbf{134}, A453 (1964).
\bibitem{williamson} S. J. Williamson, S. Foner and M. S. Dresselhaus, de Haas-van Alphen Effect in Pyrolytic and Single-Crystal Graphite, Phys. Rev. \textbf{140}, A1429 (1965).
\bibitem{slonczewski} J. C. Slonczewski and P. R. Weiss, Band Structure of Graphite, Phys. Rev. \textbf{109}, 272 (1958).
\bibitem{mcclure} J. W. McClure, Band Structure of Graphite and de Haas-van Alphen Effect, Phys. Rev. \textbf{108}, 612 (1957).
\bibitem{woollam} J. A. Woollam, Graphite Carrier Locations and Quantum Transport to 10 T (100 kG), Phys. Rev. B \textbf{3}, 1148 (1971).
\bibitem{luk} I. A. Luk'yanchuk and Y. Kopelevich, Phase Analysis of Quantum Oscillations in Graphite, Phys. Rev. Lett. \textbf{93}, 166402 (2004).
\bibitem{schneider}J. M. Schneider \emph{et al.},  Consistent Interpretation of the Low-Temperature Magnetotransport in Graphite Using the Slonczewski-Weiss-McClure 3D Band-Structure Calculations, Phys. Rev. Lett. \textbf{102}, 166403 (2009)
\bibitem{mills}J. J. Mills, R. A. Morant and D. A. Wright, Thermomagnetic effects in pyrolytic graphite, Brit. J. Appl. Phys. \textbf{16}, 479 (1965).
\bibitem{behnia3} K. Behnia, M. -A. M\'easson and Y. Kopelevich, Nernst Effect in Semimetals: The Effective Mass and the Figure of Merit, Phys. Rev. Lett. \textbf{98}, 076603 (2007)
\bibitem{sugihara} K. Sugihara and S. Ono, Galvanometric properties of graphite at low temperatures, J. Phys. Soc. Jpn. \textbf{21}, 631 (1966).
\bibitem{soule2} D. E. Soule, Magnetic Field Dependence of the Hall Effect and Magnetoresistance in Graphite Single Crystals, Phys. Rev. \textbf{112}, 698 (1958).
\bibitem{gusynin} V. P. Gusynin and S. G. Sharapov, Transport of Dirac quasiparticles in graphene: Hall and optical conductivities, Phys. Rev. B \textbf{73}, 245411 (2006).
\bibitem{fletcher} R. Fletcher,  Magnetothermoelectric effects in semiconductor systems, Semicond. Sci. Technol. \textbf{14}, R1 (1999).
\bibitem{livanov} D. V. Livanov, Hall and Nernst effects in some models of anisotropic metals, Phys. Rev. B \textbf{60}, 13439  (1999).
\bibitem{behnia2} K. Behnia, L. Balicas and Y. Kopelevich,  Signatures of electron fractionalization in ultraquantum bismuth, Science \textbf{317}, 1729 (2007).
\end{thebibliography}
\end{document}